\documentclass[runningheads]{llncs}
\usepackage{graphicx}
\usepackage{amsmath}
\usepackage{amsfonts}
\usepackage{color}
\usepackage{hyperref}
\begin{document}
\title{Low-Rank Conjugate Gradient-Net for \\ Accelerated Cardiac MR Imaging}
\titlerunning{LowRank-CGNet}
\date{\today}
%
%
\author{Jaykumar H. Patel\inst{1,2,*}\orcidID{0000-0003-1954-4879} \and \\
Brenden T. Kadota\inst{1,2}\orcidID{0009-0001-1510-7746} \and \\
Calder D. Sheagren\inst{1,2}\orcidID{0000-0001-7439-2906} \and \\
Mark Chiew\inst{1,2}\orcidID{0000-0001-6272-8783} \and \\
Graham A. Wright\inst{1,2}}
\authorrunning{J. Patel et al.}
%
\institute{$^1$Department of Medical Biophysics, University of Toronto, Toronto, Ontario, Canada 
\\ $^2$Physical Sciences Platform, Sunnybrook Research Institute, Ontario, Canada
*Corresponding email: \email{jaykumar.patel@mail.utoronto.ca}}
\maketitle              
\begin{abstract}
Cardiovascular diseases (CVDs) remain the leading cause of mortality and morbidity worldwide. Both diagnosis and prognosis of these diseases benefit from high-quality imaging, which cardiac magnetic resonance imaging provides. CMR imaging requires lengthy acquisition times and multiple breath-holds for a complete exam, which can lead to patient discomfort and frequently results in image artifacts. In this work, we present a Low-rank tensor U-Net method (LowRank-CGNet) that rapidly reconstructs highly undersampled data with a variety of anatomy, contrast, and undersampling artifacts. The model uses conjugate gradient data consistency to solve for the spatial and temporal bases and employs a U-Net to further regularize the basis vectors. Currently, model performance is superior to a standard U-Net, but inferior to conventional compressed sensing methods. In the future, we aim to further improve model performance by increasing the U-Net size, extending the training duration, and dynamically updating the tensor rank for different anatomies.
\keywords{Multi-contrast Imaging, k-t imaging, Deep subspace learning, Low-Rank Tensor Model, U-Net and Unrolled Model}
\end{abstract}
\section{Introduction}

Cardiovascular diseases (CVDs) remain the leading cause of mortality and morbidity worldwide, necessitating early diagnosis, disease monitoring, and prognosis assessment post-treatment, all involving imaging \cite{tseng2016introduction}. Cardiac magnetic resonance (CMR) imaging, a non-invasive modality devoid of ionizing radiation, is often employed for diagnosing and managing CVDs. CMR imaging utilizes various tissue contrast mechanisms to provide detailed and comprehensive assessments of the heart's structure and function for a range of cardiac pathologies and morphologies.

CMR is a powerful tool that offers dynamic imaging at high spatiotemporal resolution and varying image contrast for accurate diagnosis. T1 and T2 mapping are crucial in CMR imaging due to their ability to provide quantitative assessments of myocardial tissue properties. T1 mapping evaluates extracellular volume and diffuse myocardial fibrosis for detecting cardiomyopathies \cite{henningsson_black-blood_2022,perea2015t1}, while T2 mapping identifies edematous regions within myocardial tissue. Cine imaging, with its high temporal resolution, captures cardiac motion, allowing the measurement of ventricular volumes and ejection fraction. Cine imaging is the gold standard for assessing ventricular function and plays a pivotal role in determining cardiac health~\cite{menchon2019reconstruction}. Tagging techniques also provide functional information regarding myocardial strain and deformation \cite{axel1989mr}. Additionally, imaging of the aorta using CMR is vital for assessing aortic pathologies such as aneurysms and dissections \cite{stankovic20144d}. These advanced CMR techniques enhance diagnostic accuracy by providing detailed spatiotemporal information about cardiac and vascular structures and functions.

However, a significant drawback of these techniques is the extended acquisition time, which can lead to patient discomfort and frequently results in image artifacts from cardiac, respiratory, and patient motion, compromising spatial resolution and signal-to-noise ratio (SNR) \cite{sheagren_motion-compensated_2023}. This necessitates re-scans, the use of motion compensation or correction methods, and the use of triggering techniques to mitigate artifacts and/or accelerated  undersampled scans to reduce the acquisition duration. These limitations diminish patient throughput and accessibility. To shorten image acquisition times, parallel imaging techniques such as Sensitivity Encoding (SENSE) and Generalized Autocalibrating Partial Parallel Acquisition (GRAPPA) are employed to exploit multi-channel spatial redundancy and accelerate scans by acquiring a fraction of the fully-sampled data \cite{pruessmann_sense_1999,griswold_generalized_2002}. 
Dynamic undersampled datasets can be reconstructed using methods that exploit spatiotemporal ($k$-$t$) redundancy, including compressed sensing and low-rank approaches. While compressed sensing techniques are effective and can leverage sparsity in the $k$-$t$ domain, they often require extensive empirical parameter tuning tailored to each patient and imaging contrast, which can reduce generalizability. 

In contrast, low-rank tensor formulations and subspace regularization methods provide a more flexible model, where correlations between global and local features are exploited across spatiotemporal domains. This flexibility enables low-rank methods to deliver superior reconstruction quality and consistency, making them a more robust choice across diverse imaging contrasts~\cite{tanner2023compressed}. These approaches improve the overall image quality of reconstructions by addressing varied undersampled spatial patterns that arise from cardiac motion. For instance, spatiotemporal images for T1/T2 mapping, cine imaging, and aorta imaging, which are composed of spatial images acquired over different temporal domains (inversion times, echo times, cardiac phases, or respiratory phases), benefit significantly from these techniques~\cite{yaman2019low,ma2020dynamic}.

Low-rank tensor formulations can be employed to set singular vectors with lower-magnitude singular values - corresponding to noise and artifacts - to zero \cite{cao2024alternating}. This process enhances image quality by filtering out undesirable elements. Additionally, subspace regularization can further refine dynamic image quality. Low-rank algorithms are particularly advantageous due to their ability to compactly represent a wide range of signal profiles, making them highly suitable for generalizing across different imaging contrasts \cite{sheagren2023accelerated}.

To generalize over the cardiac dynamic contrast subspace, separate spatial and temporal regularizers can also be added to improve image quality and preserve dynamic spatial information across generalized image contrasts. In this paper, we propose a temporally dynamic generalized MRI reconstruction method using a subspace low-rank model, which aims to recover spatiotemporal bases with the U-Net backbone of the generalized accelerated dynamic cardiac images.

\section{Methods}

\subsection{Dataset}

Briefly, this method was trained on the CMRxRecon2024 dataset \cite{wang_cmrxrecon2024}. The dataset consists of 2D Cartesian-sampled MRI images of the heart and aorta. A prospective acceleration factor of 2 was used to acquire the data, with GRAPPA reconstruction being used to create a fully sampled k-space. Subsequently, data was undersampled using various $k$-$t$ methods, which include uniform undersampling, Gaussian-density undersampling, and pseudo-radial undersampling. Each pattern preserved a fully sampled central k-space region for coil sensitivity calculation. 

\subsection{Algorithm}

\begin{figure}[t]
    \centering
    \includegraphics[width=0.6\textwidth]{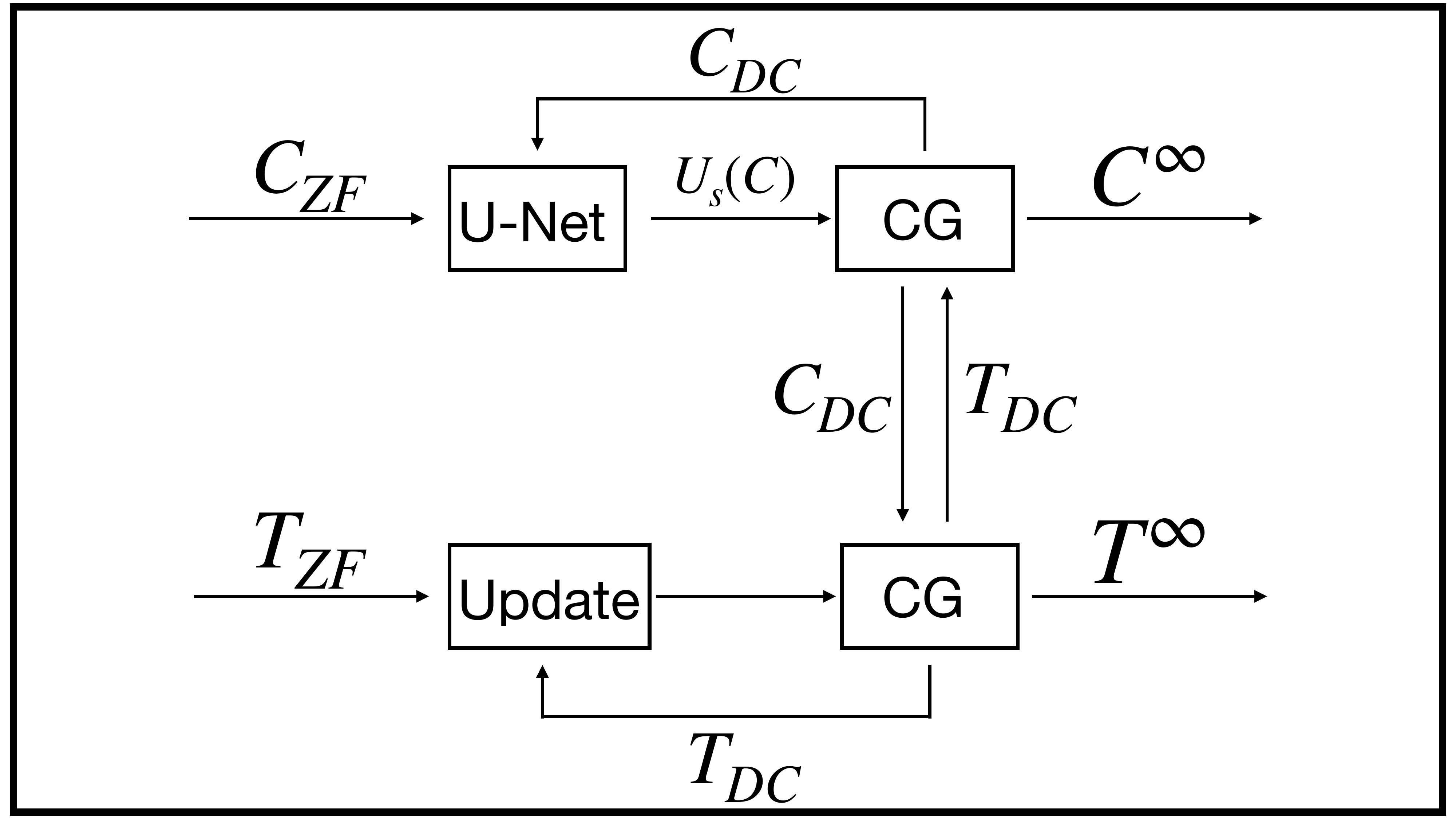}
    \includegraphics[width=0.6\textwidth]{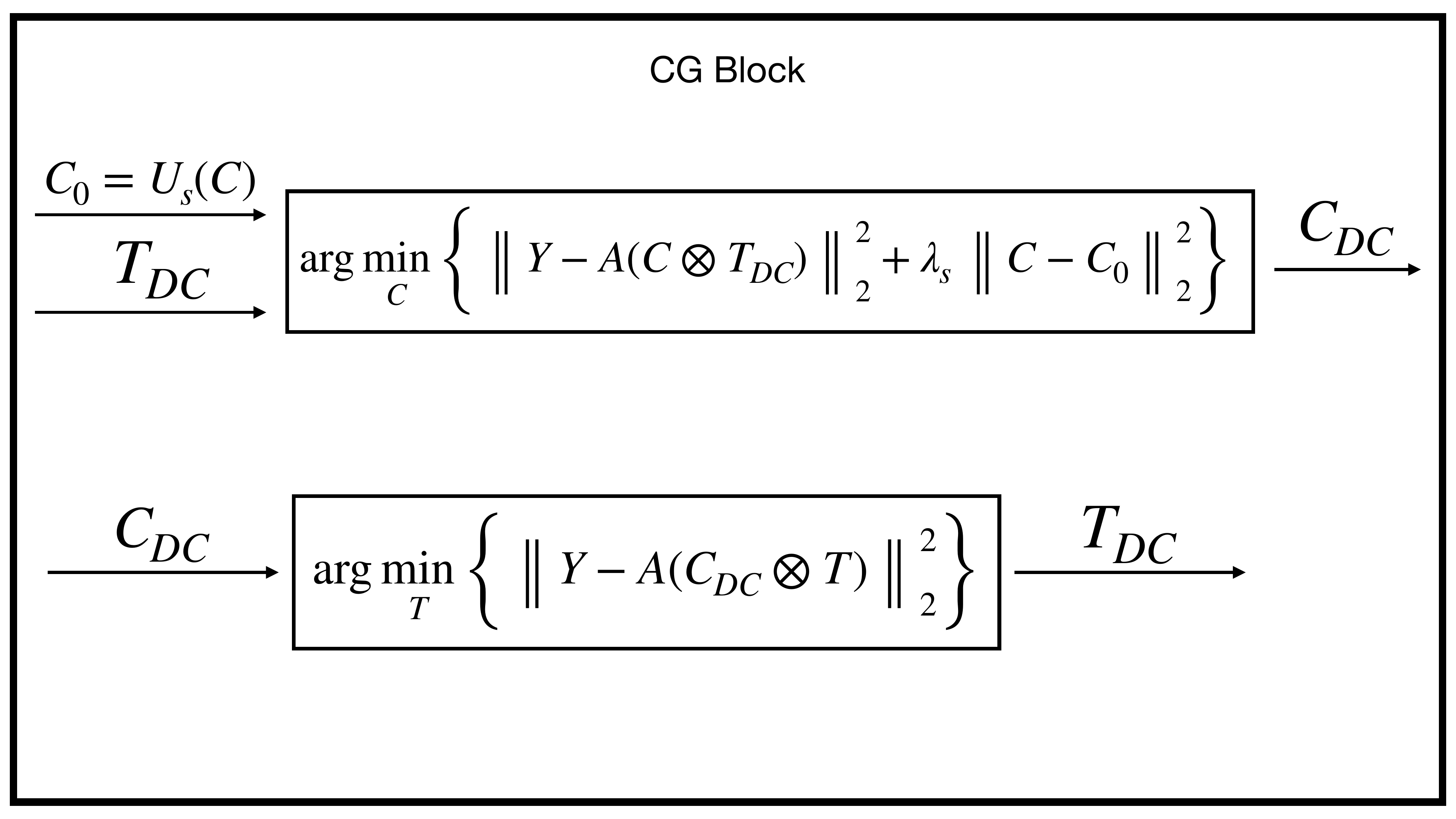}
    \caption{Visual overview of algorithm. Initial spatial and temporal bases $C_{\rm ZF}$ and $T_{\rm ZF}$ are fed into the model. Spatial bases are fed through a U-Net, before a conjugate gradient-based update step is used to generate data-consistent spatial bases. Subsequently, a separate conjugate gradient update step is used to generate data-consistent temporal bases. These steps are iterated until convergence.}
    \label{fig::algorithm}
\end{figure}

Define $X$ to be the image we want to reconstruct, with $X_q, q = 1, \ldots, Q$ being images from individual coils. We define our forward model as 
\begin{equation}
   A: \mathbb{C}^{N_x \times N_y} \to \mathbb{C}^{N_x \times N_y \times N_c}; \quad X \mapsto (PFS_1 X, PFS_2 X, \ldots, PFS_QX).
\end{equation}
Here, $S_q$ represents multiplication by the sensitivity maps for coil $q$; $F$ represents the Fourier Transform; and $P$ represents the projection onto the sampled points. Similarly, we can define $\tilde P$ to be the projection onto the points not sampled by $P$ and define a corresponding operator 
\begin{equation}
   \tilde A: \mathbb{C}^{N_x \times N_y} \to \mathbb{C}^{N_x \times N_y \times N_c}; \quad X \mapsto (\tilde PFS_1 X, \tilde PFS_2 X, \ldots, \tilde PFS_QX).
\end{equation}
We denote the k-space data by $Y \in \mathbb{C}^{N_x \times N_y \times N_z}$, with the relation $AX = Y$. 

In this reconstruction method, we decompose $X = C\otimes T$ to be a tensor product of a spatial basis $C$ and a temporal basis $T$. We aim to jointly solve for data consistency of $X$ and spatial regularization of $C$, denoted as $R_s(C)$,
\begin{equation}
    \hat C, \hat T = \arg\min_{C, T}\left\{{1\over 2}\left\|Y - A(C\otimes T)\right\|_2^2 + \lambda R_s(C)\right\}.
\end{equation}
We aim to solve this problem via an alternating minimization of conjugate gradient data consistency on $C$ and $T$ separately, with a trained model $U_s$ that provides regularization of spatial bases at each step of the reconstruction: 
\begin{align}
    \hat C^k &= \arg\min_C\left\{\left\| Y - A(C \otimes T^{k-1})\right\|_2^2 + \lambda\left\|C - U_s(C^{k-1})\right\|_2^2\right\} \\
    \hat T^k &= \arg\min_T\left\{\left\| Y - A(C^k \otimes T)\right\|_2^2\right\}.
\end{align}

As a final step, to ensure consistency with acquired data points, we set 
\begin{equation}
    X = A^H(Y + \tilde A (C^\infty \otimes T^\infty)).
\end{equation}

\subsection{Model Training}\label{sec::model-training}
We train our Low-rank tensor U-Net model (LowRank-CGNet) in an end-to-end manner with a U-Net spatial denoiser and conjugate gradient data consistency steps. The U-Net~\cite{ronneberger_u-net_2015} has an initial filter size of 32 with 4 down-sampling layers. We make a modification to the original U-Net by adding instance norm after each convolution which has been shown to improve texture recovery \cite{ulyanov_improved_2017}. The initial estimate for the temporal basis was derived from the 16x16 centre box of k-space and truncated to a rank of 3. The initial estimate of the spatial basis was determined by solving Equation 4 with $\lambda=0$ using conjugate gradient for 4 steps. The initial estimates were then passed through the unrolled architecture with alternating steps of conjugate gradient data consistency and spatial denoising. The U-Net denoisers were trained to map to the residuals
\begin{equation}
    \hat{C}_{k+1} = C_k + U_s(C_k)
\end{equation}
where $C_k$ is the input estimated bases at unrolled step $k$ and $\hat{C}_{k+1}$, are the final denoised bases. Each basis was passed independently through the network with the complex and real data split along the channel dimension.

A mixed $\ell^1$ and SSIM loss is used defined by 
\begin{equation}
L(\hat X, X) = \frac{\lVert\hat{X} - X\rVert_1}{\lVert\hat{X}\rVert_1} + \beta {\rm SSIM}(\hat{X}, X)
\end{equation} 
with $\beta = 10^{-1}$. A learning rate of $10^{-3}$ was used with an ADAM optimizer. The training was done on all slices, with a new undersampling mask randomly selected for each slice every epoch. The training was stopped after 20 epochs on 4$\times$ NVIDIA P100 Pascal (16G HBM2 memory) which took around 48 hours. All models were implemented in PyTorch and trained using PyTorch Lightning.  Additionally, we trained a comparison U-Net model directly on aliased zero-filled images with the same training parameters as above.

K-space data was pre-processed by normalizing the maximum k-space point to 1 and zero padding of $k_x$ and $k_y$ to [512, 256]. A batch size of 4 was used for both models and the internal validation was done on 10$\%$ of the training dataset. 

\subsection{Experiments}
Model performance in undersampled images was compared to the fully sampled reference images, within the validation set. Peak signal-to-noise ratio (PSNR), structural similarity index (SSIM), and normalized mean squared error (NMSE) metrics were used. The results for our new method were benchmarked against conventional parallel imaging reconstruction methods, including CG-SENSE, $\ell^1$-ESPIRiT, and U-Net, all as scored by the Synapse validation portal. 


\section{Results}\label{sec::result}

\begin{figure}
    \centering
    \includegraphics[width=0.8\linewidth]{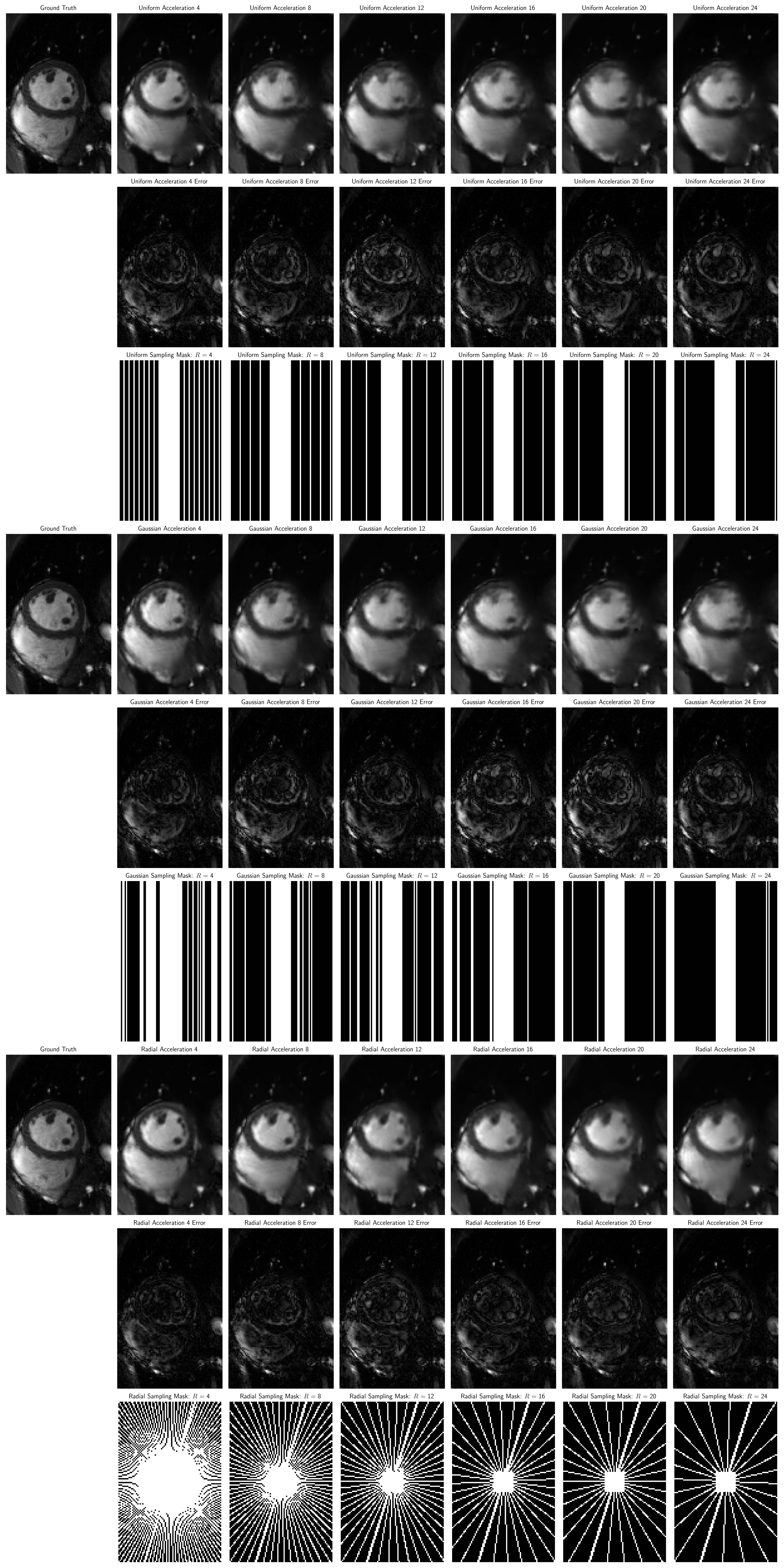}
    \caption{Comparison of reconstructed Short-Axis CINE MRI images at varying acceleration rates using Uniform, Gaussian, and Radial sampling patterns. Each row displays images at different acceleration factors (R = 4, 8, 12, 16, 20, 24), showing the reconstructed images, corresponding error maps, and sampling masks for each technique. The first column presents the Ground Truth for reference. As the acceleration rate increases, image quality degradation is observed across all sampling patterns, with Radial sampling showing relatively less error at higher acceleration rates compared to Uniform and Gaussian sampling.}
    \label{fig::acceleration-comparison}
\end{figure}

\begin{table}[t]
    \centering
    \begin{tabular}{|l|l|l|l|l|l|}
    \hline
    Metric  & Zero-Filled & CG-SENSE & $\ell^1$-ESPIRiT & U-Net & Proposed  \\
    \hline
     SSIM [1] & 0.46 & 0.56 & \textbf{0.67} & 0.56 & 0.66\\
    \hline
     PSNR [dB] & 19.42 & 21.92 & 25.40 & 23.25 & \textbf{30.06}\\
    \hline
     NMSE [1] & 0.29 & 0.21 & \textbf{0.15} & 0.22 & 0.31\\
    \hline
    \end{tabular}
    \caption{CMRxRecon Validation dataset results for LowRank-CGNet and reference reconstruction methods, as scored by the online Synapse portal.}
    \label{tab::validation-results}
\end{table}

LowRank-CGNet was successfully implemented, and a model was trained to convergence. The model's performance on the internal validation set during the training was SSIM: 0.91, PSNR: 35.5 and NMSE: 0.066. Comparing quantitative metrics on the CMRxRecon validation set, LowRank-CGNet slightly outperforms the conventional U-Net, but does not perform at the level of $\ell^1$-ESPIRiT (Table \ref{tab::validation-results}). This may be due to the limited generalization capacity of the model when compared to compressed sensing, which requires several tunable parameters. Further, $\ell^1$-ESPIRiT is an iterative algorithm that requires more computational time than LowRank-CGNet, which may be infeasible for the time constraints of real-world clinical applications. Our model performed similarly on the CMRxRecon testing set, with PSNR = 30.6dB, SSIM = 0.69, NMSE = 0.31. 

\begin{figure}[t]
    \centering
    \includegraphics[width=1\linewidth]{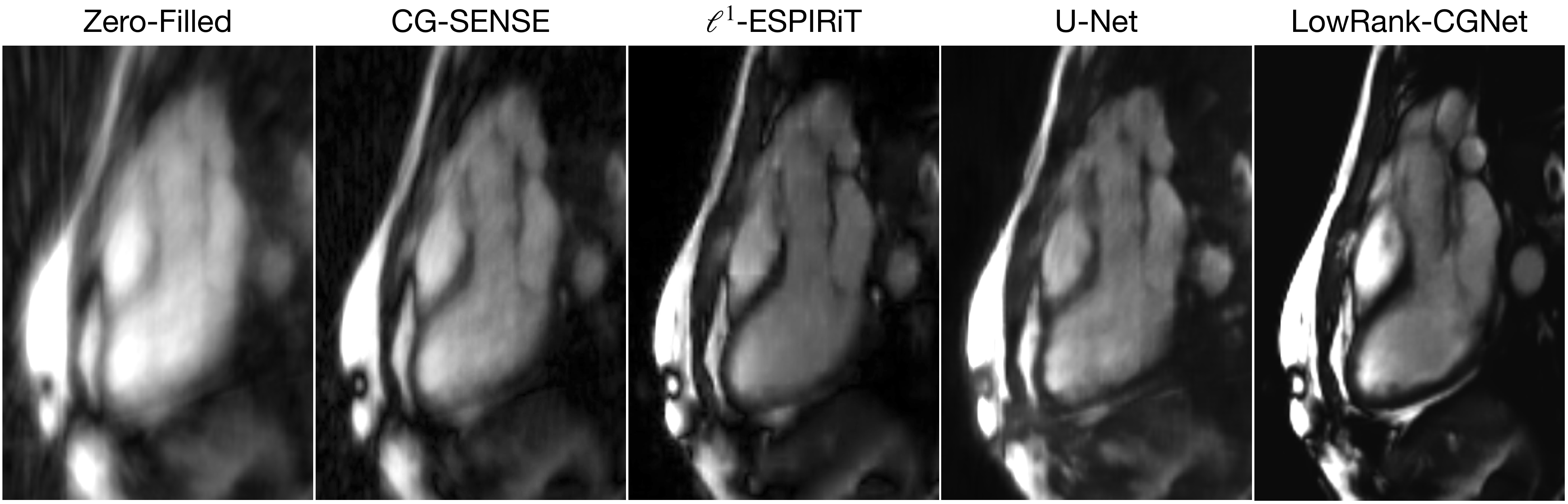}
    \caption{Comparison of reconstruction methods on a sample cine long-axis dataset, acquired with a radial undersampling pattern and 20-fold acceleration. Columns represent different reconstruction techniques.}
    \label{fig::method-comparison}
\end{figure}

Visually, LowRank-CGNet images appear natural and well spatially encoded. Figure \ref{fig::method-comparison} contains sample 20-fold accelerated radial cine long-axis images from multiple reconstruction techniques. Zero-filled images have blurring and streaking artifacts, which are marginally corrected by CG-SENSE. $\ell^1$-ESPIRiT images are blurrier, with some unnatural block-artifacts arising from the compression. U-Net images appear reasonably sharper compared to $\ell^1$-ESPIRiT, but have significant artifacts in the blood pool signal. LowRank-CGNet images look visually similar to $\ell^1$-ESPIRiT, with more complex blood-pool signal variations than $\ell^1$-ESPIRiT. When comparing our model performance across sampling patterns and acceleration factors, we observe increasing blur artifacts as the acceleration factor increases, which is most prominent on the uniform and Gaussian sampling pattern (Figure \ref{fig::acceleration-comparison}). 

\begin{figure}[t]
    \centering
    \includegraphics[width=\linewidth]{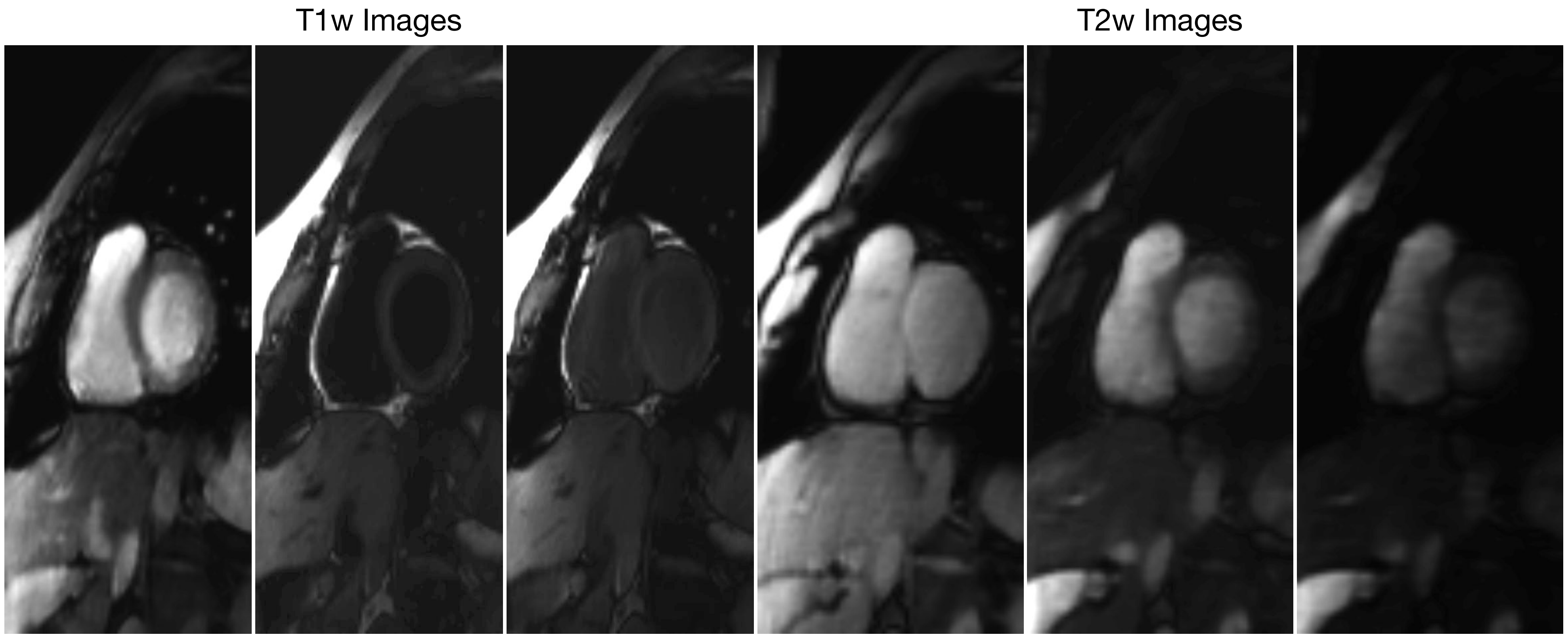}
    \caption{Sample LowRank-CGNet reconstructions for T1-weighted (T1w) and T2-weighted (T2w) images from the mapping dataset, acquired with an eight-fold uniform undersampling pattern. Left three columns: T1w images. Right three columns: T2w images. }
    \label{fig::mapping}
\end{figure}

\begin{figure}
    \centering
    \includegraphics[width=1\linewidth]{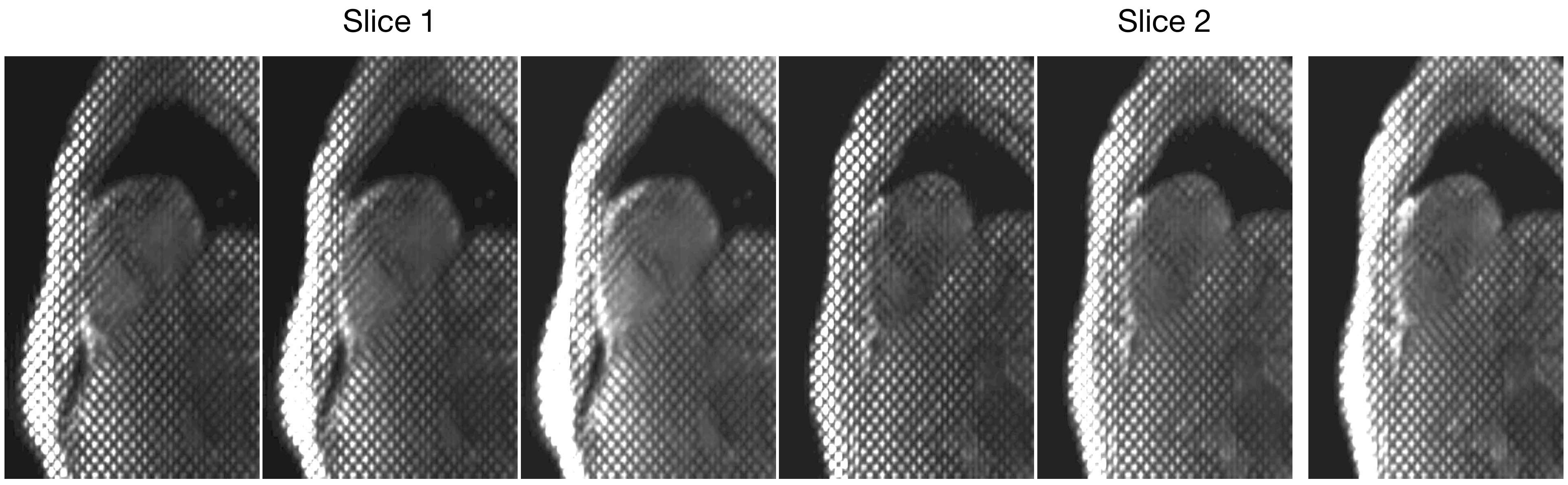}
    \caption{Sample tagging images for three time-frames in two representative slices acquired with a 20-fold radial undersampling pattern. Images are generally free from aliasing artifacts, and tag lines are visible, as expected. }
    \label{fig::tagging}
\end{figure}

\begin{figure}
    \centering
    \includegraphics[width=1\linewidth]{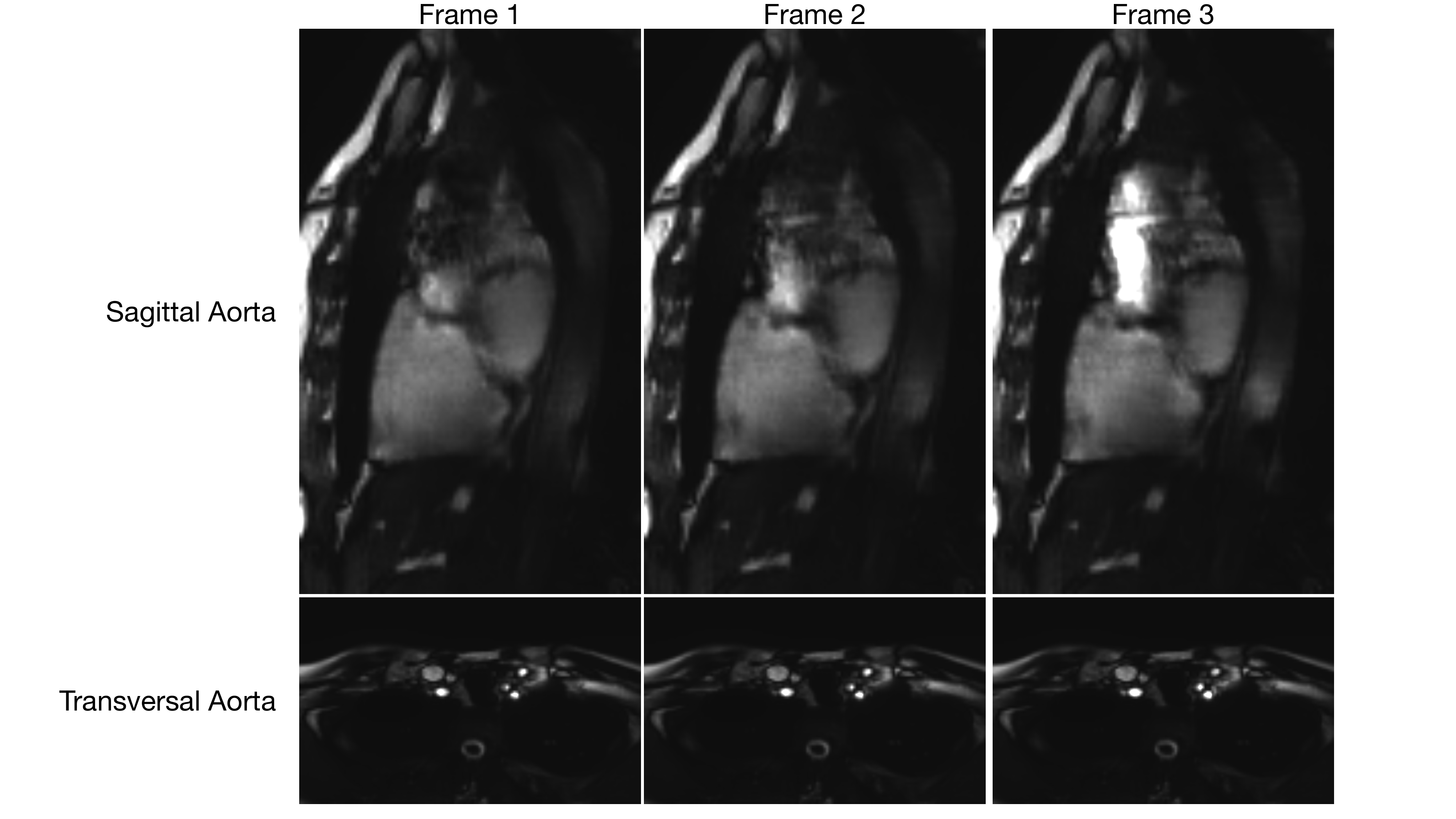}
    \caption{Sample time-frames for one representative slice of transversal and sagittal aorta images, acquired with a 24-fold Gaussian undersampling pattern. Transversal images have reasonable image quality, whereas sagittal images are corrupted by flow artifacts. Note the changes in blood flow in the sagittal images, where the region of hypointensity travels up the ascending aorta from the leftmost image to the rightmost image, and a hyperintensity appears just past the aortic valve.}
    \label{fig::aorta}
\end{figure}

Figure \ref{fig::mapping} shows sample T1-weighted and T2-weighted images from the mapping sequence, acquired with an 8-fold accelerated uniform undersampling pattern. Magnetization recovery dynamics are present, but spatial blurring and aliasing artifacts remain. Figure \ref{fig::tagging} shows sample tagging images from two representative slices of a 20-fold accelerated radial acquisition. Images appear visually free of spatial aliasing, and the tag lines appear as expected, encoding myocardial motion. Figure \ref{fig::aorta} shows sample transversal and sagittal aorta images, acquired with a 24-fold Gaussian-density undersampling pattern. Transversal images have good image quality, free of aliasing artifacts, whereas the sagittal images are affected by flow artifacts which compromises the reconstruction quality. Regions of high signal such as the right ventricle appear visually accurate, but regions of low and intermediate signal such as the ascending aortic arch have reduced image quality.

\section{Discussion}
Overall, we have developed and implemented a Low-rank tensor U-Net model, LowRank-CGNet, that rapidly reconstructs highly undersampled data with a variety of anatomy, contrast, and undersampling artifacts. Image quality is generally adequate, with variations in the degree of undersampling artifacts across anatomies and acceleration factors. Generally, LowRank-CGNet performed the worst on uniformly accelerated data and better on Gaussian-density and radial data. This agrees with prior literature that temporal incoherence enables highly accelerated reconstruction with minimal artifacts \cite{feng_golden-angle_2014}. Further, residual artifacts in radial undersampling are more benign than the one-dimensional Cartesian undersampling, due to the oversampling of low-order k-space. 

Our method has several limitations. First, the low-rank tensor framework requires a tunable parameter, namely the rank of the tensor. The optimal rank may vary across modalities, since mapping images have three to nine temporal frames, whereas cine images can have 10 to 15 or more temporal frames. As such, we will investigate changing the tensor rank based on the anatomy type. Second, residual aliasing artifacts in data such as eight-fold accelerated mapping indicate that the model may not be sufficiently regularizing the data. As such, we will look to increase the number of model parameters to generalize better to unknown anatomy. While LowRank-CGNet images appear sharper with fewer spatial artifacts compared to $\ell^1$-ESPIRiT, inadequate data scaling may reduce the quantitative scores, affecting model performance.

A final limitation of this work is that the reference data was prospectively undersampled by a factor of 2, with parallel imaging used to estimate the missing k-space. As shown by Frenklak et al, increasing prospective acceleration factors causes overly optimistic model performance within the given dataset, which leads to biased downstream perception of algorithm performance \cite{frenklak_ismrm_2024}. A potential extension of the challenge could be recovering the two-fold undersampled k-space data directly from subsampled measurements of the acquired k-space lines. 

\section{Conclusion}
In conclusion, we have developed and implemented a Low-rank tensor U-Net method, LowRank-CGNet, that rapidly reconstructs highly undersampled data with a variety of anatomy, contrast, and undersampling artifacts. In future, we hope to further improve the model performance by increasing the U-Net size and dynamically updating the tensor rank for different anatomies. 

\section*{Acknowledgments}

This work utilized the computing resources of the Digital Research Alliance of Canada. JP, CS, and GW received funding from the Canadian Institutes of Health Research (PJT178299), while BK and MC received support from the Canada Research Chairs Program and the NSERC Discovery Grant (RGPIN/2023-03410).


\end{document}